
\input harvmac.tex



\def\unlockat{\catcode`\@=11}
\def\lockat{\catcode`\@=12}

\unlockat

\def\newsec#1{\global\advance\secno by1\message{(\the\secno. #1)}
\global\subsecno=0\global\subsubsecno=0\eqnres@t\noindent
{\bf\the\secno. #1}
\writetoca{{\secsym} {#1}}\par\nobreak\medskip\nobreak}
\global\newcount\subsecno \global\subsecno=0
\def\subsec#1{\global\advance\subsecno
by1\message{(\secsym\the\subsecno. #1)}
\ifnum\lastpenalty>9000\else\bigbreak\fi\global\subsubsecno=0
\noindent{\it\secsym\the\subsecno. #1}
\writetoca{\string\quad {\secsym\the\subsecno.} {#1}}
\par\nobreak\medskip\nobreak}
\global\newcount\subsubsecno \global\subsubsecno=0
\def\subsubsec#1{\global\advance\subsubsecno by1
\message{(\secsym\the\subsecno.\the\subsubsecno. #1)}
\ifnum\lastpenalty>9000\else\bigbreak\fi
\noindent\quad{\secsym\the\subsecno.\the\subsubsecno.}{#1}
\writetoca{\string\qquad{\secsym\the\subsecno.\the\subsubsecno.}{#1}}
\par\nobreak\medskip\nobreak}

\def\subsubseclab#1{\DefWarn#1\xdef
#1{\noexpand\hyperref{}{subsubsection}%
{\secsym\the\subsecno.\the\subsubsecno}%
{\secsym\the\subsecno.\the\subsubsecno}}%
\writedef{#1\leftbracket#1}\wrlabeL{#1=#1}}
\lockat

\def\IL{\relax{\rm I\kern-.18em L}}
\def\IH{\relax{\rm I\kern-.18em H}}
\def\IR{\relax{\rm I\kern-.18em R}}
\def\IC{\relax\hbox{$\inbar\kern-.3em{\rm C}$}}
\def\IZ{\relax\ifmmode\mathchoice
{\hbox{\cmss Z\kern-.4em Z}}{\hbox{\cmss Z\kern-.4em Z}}
{\lower.9pt\hbox{\cmsss Z\kern-.4em Z}}
{\lower1.2pt\hbox{\cmsss Z\kern-.4em Z}}\else{\cmss Z\kern-.4em
Z}\fi}
\def\CM {{\cal M}}

\def\CR {{\cal R}}

\def\CP {{\cal P }}
\def\CL {{\cal L}}
\def\CV {{\cal V}}
\def\CO {{\cal O}}
\def\CZ {{\cal Z}}
\def\CE {{\cal E}}
\def\CG {{\cal G}}
\def\CH {{\cal H}}

\def\CB {{\cal B}}
\def\CS {{\cal S}}
\def\CA{{\cal A}}
\def\gpg{g^{-1} \p g}

\def\CM {{\cal M}}

\def\CO {{\cal O}}

\def\CP {{\cal P }}

\def\CE{{\cal E }}
\def\CV{{\cal V }}
\def\CZ {{\cal Z }}
\def\CS {{\cal S }}
\def\ch{{\rm ch}}
\def\CY{{\cal Y }}

\def\gpg{g^{-1} \p g}

\def\zb {\bar{z}}

\font\manual=manfnt \def\dbend{\lower3.5pt\hbox{\manual\char127}}

\def\IZ{\relax\ifmmode\mathchoice
{\hbox{\cmss Z\kern-.4em Z}}{\hbox{\cmss Z\kern-.4em Z}}
{\lower.9pt\hbox{\cmsss Z\kern-.4em Z}}
{\lower1.2pt\hbox{\cmsss Z\kern-.4em Z}}\else{\cmss Z\kern-.4em
Z}\fi}
\def\half {{1\over 2}}

\def\p{\partial}
\def\pb{\bar{\partial}}

\def\CM {{\cal M}}

\def\CO {{\cal O}}

\def\CP {{\cal P }}

\def\CE{{\cal E }}
\def\CV{{\cal V }}
\def\CZ {{\cal Z }}
\def\CS {{\cal S }}
\def\ch{{\rm ch}}

\def\DET{{\rm DET}}

\def\gpg{g^{-1} \p g}


\def\IZ{\relax\ifmmode\mathchoice
{\hbox{\cmss Z\kern-.4em Z}}{\hbox{\cmss Z\kern-.4em Z}}
{\lower.9pt\hbox{\cmsss Z\kern-.4em Z}}
{\lower1.2pt\hbox{\cmsss Z\kern-.4em Z}}\else{\cmss Z\kern-.4em
Z}\fi}
\def\IB{\relax{\rm I\kern-.18em B}}
\def\IC{{\relax\hbox{$\inbar\kern-.3em{\rm C}$}}}
\def\ID{\relax{\rm I\kern-.18em D}}
\def\IE{\relax{\rm I\kern-.18em E}}
\def\IF{\relax{\rm I\kern-.18em F}}
\def\IG{\relax\hbox{$\inbar\kern-.3em{\rm G}$}}
\def\IGa{\relax\hbox{${\rm I}\kern-.18em\Gamma$}}
\def\IH{\relax{\rm I\kern-.18em H}}
\def\II{\relax{\rm I\kern-.18em I}}
\def\IK{\relax{\rm I\kern-.18em K}}
\def\IP{\relax{\rm I\kern-.18em P}}

\def\jb{{\bar j}}
\def\lieg{{\underline{\bf g}}}
\def\liet{{\underline{\bf t}}}

\def\inbar{\,\vrule height1.5ex width.4pt depth0pt}
\def\p{\partial}
\def\pb{{\bar \p}}

\font\cmss=cmss10 \font\cmsss=cmss10 at 7pt
\def\IR{\relax{\rm I\kern-.18em R}}

\def\Tr{\rm Tr}
\def\vol{{\rm vol}}
\def\Vol{{\rm Vol}}

\def\wzwt{$WZW_2$}
\def\wzwf{$WZW_4$}

\def\zb {{\bar{z}}}

\def\boxit#1{\vbox{\hrule\hbox{\vrule\kern8pt
\vbox{\hbox{\kern8pt}\hbox{\vbox{#1}}\hbox{\kern8pt}}
\kern8pt\vrule}\hrule}}
\def\mathboxit#1{\vbox{\hrule\hbox{\vrule\kern8pt\vbox{\kern8pt
\hbox{$\displaystyle #1$}\kern8pt}\kern8pt\vrule}\hrule}}


\def\jb{{\bar j}}

\def\lieg{{\underline{\bf g}}}

\def\liet{{\underline{\bf t}}}

\def\inbar{\,\vrule height1.5ex width.4pt depth0pt}

\def\p{\partial}

\def\pb{{\bar \p}}

\font\cmss=cmss10 \font\cmsss=cmss10 at 7pt
\def\IR{\relax{\rm I\kern-.18em R}}

\def\Tr{\rm Tr}

\def\vol{{\rm vol}}
\def\Vol{{\rm Vol}}

\def\wzwt{$WZW_2$}
\def\wzwf{$WZW_4$}

\def\zb {{\bar{z}}}


\def\a1{{\cal A}^{1,1}}

\def\bpsi{\bar \psi}
\def\hi{\chi^{2,0}}
\def\bhi{\chi^{0,2}}
\def\hh{H^{2,0}}
\def\bh{H^{0,2}}

\def\pa{\p_{A}}
\def\pba{\pb_{\bar A}}


%
\lref\blzh{A. Belavin, V. Zakharov, ``Yang-Mills Equations as inverse
scattering
problem''Phys. Lett. B73, (1978) 53}
\lref\afs{Alekseev, Faddeev, Shatashvili,  }
\lref\bost{L. Alvarez-Gaume, J.B. Bost , G. Moore, P. Nelson, C.
Vafa,
``Bosonization on higher genus Riemann surfaces,''
Commun.Math.Phys.112:503,1987}
\lref\agmv{L. Alvarez-Gaum\'e,
C. Gomez, G. Moore,
and C. Vafa, ``Strings in the Operator Formalism,''
Nucl. Phys. {\bf 303}(1988)455}
\lref\atiyah{M. Atiyah, ``Green's Functions for
Self-Dual Four-Manifolds,'' Adv. Math. Suppl.
{\bf 7A} (1981)129}

\lref\AHS{M.~ Atiyah, N.~ Hitchin and I.~ Singer, ``Self-Duality in
Four-Dimensional
Riemannian Geometry", Proc. Royal Soc. (London) {\bf A362} (1978)
425-461.}
\lref\fmlies{M. F. Atiyah and I. M. Singer,
``The index of elliptic operators IV,'' Ann. Math. {\bf 93}(1968)119}
\lref\bagger{E. Witten and J. Bagger, Phys. Lett.
{\bf 115B}(1982)202}
\lref\banks{T. Banks, ``Vertex Operators in 2K Dimensions,''
hep-th/9503145   }
\lref\berk{N. Berkovits, ``Super-Poincare Invariant Superstring Field
Theory''
hep-th/9503099 }
\lref\biquard{O. Biquard, ``Sur les fibr\'es paraboliques
sur une surface complexe,'' to appear in J. Lond. Math.
Soc.}
\lref\bjsv{hep-th/9501096,
Topological Reduction of 4D SYM to 2D $\sigma$--Models,
 M. Bershadsky, A. Johansen, V. Sadov and C. Vafa }
\lref\BlThlgt{M.~ Blau and G.~ Thompson, ``Lectures on 2d Gauge
Theories: Topological Aspects and Path
Integral Techniques", Presented at the
Summer School in Hogh Energy Physics and
Cosmology, Trieste, Italy, 14 Jun - 30 Jul
1993, hep-th/9310144.}
\lref\bpz{A.A. Belavin, A.M. Polyakov, A.B. Zamolodchikov,
``Infinite conformal symmetry in two-dimensional quantum
field theory,'' Nucl.Phys.B241:333,1984}
\lref\braam{P.J. Braam, A. Maciocia, and A. Todorov,
``Instanton moduli as a novel map from tori to
K3-surfaces,'' Inven. Math. {\bf 108} (1992) 419}
\lref\cllnhrvy{Callan and Harvey, Nucl
Phys. {\bf B250}(1985)427}
\lref\CMR{ For a review, see
S. Cordes, G. Moore, and S. Ramgoolam,
`` Lectures on 2D Yang Mills theory, Equivariant
Cohomology, and Topological String Theory,''
Lectures presented at the 1994 Les Houches Summer School
 ``Fluctuating Geometries in Statistical Mechanics and Field
Theory.''
and at the Trieste 1994 Spring school on superstrings.
hep-th/9411210, or see http://xxx.lanl.gov/lh94}
\lref\devchand{Ch. Devchand and V. Ogievetsky,
``Four dimensional integrable theories,'' hep-th/9410147}
\lref\devchandi{
Ch. Devchand and A.N. Leznov,
``B \"acklund transformation for supersymmetric self-dual theories
for
semisimple
gauge groups and a hierarchy of $A_1$ solutions,'' hep-th/9301098,
Commun. Math. Phys. {\bf 160} (1994) 551}
\lref\dnld{S. Donaldson, ``Anti self-dual Yang-Mills
connections over complex  algebraic surfaces and stable
vector bundles,'' Proc. Lond. Math. Soc,
{\bf 50} (1985)1}

\lref\DoKro{S.K.~ Donaldson and P.B.~ Kronheimer,
{\it The Geometry of Four-Manifolds},
Clarendon Press, Oxford, 1990.}
\lref\donii{
S. Donaldson, Duke Math. J. , {\bf 54} (1987) 231. }

\lref\elitzur{S. Elitzur, G. Moore,
A. Schwimmer, and N. Seiberg,
``Remarks on the Canonical Quantization of the Chern-Simons-
Witten Theory,'' Nucl. Phys. {\bf B326}(1989)108;
G. Moore and N. Seiberg,
``Lectures on Rational Conformal Field Theory,''
, in {\it Strings'89},Proceedings
of the Trieste Spring School on Superstrings,
3-14 April 1989, M. Green, et. al. Eds. World
Scientific, 1990}
\lref\etingof{P.I. Etingof and I.B. Frenkel,
``Central Extensions of Current Groups in
Two Dimensions,'' Commun. Math.
Phys. {\bf 165}(1994) 429}

\lref\evans{M. Evans, F. G\"ursey, V. Ogievetsky,
``From 2D conformal to 4D self-dual theories:
Quaternionic analyticity,''
Phys. Rev. {\bf D47}(1993)3496}
\lref\fs{L. Faddeev and S. Shatashvili, Theor. Math. Fiz., 60 (1984)
206}
\lref\fsi{ L. Faddeev, Phys. Lett. B145 (1984) 81.}
\lref\fz{I. Frenkel, I. Singer, unpublished.}
\lref\fk{I. Frenkel and B. Khesin, ``Four dimensional
realization of two dimensional current groups,'' Yale
preprint, July 1995, to appear in Commun. Math. Phys.}
\lref\galperin{A. Galperin, E. Ivanov, V. Ogievetsky,
E. Sokatchev, Ann. Phys. {\bf 185}(1988) 1}
\lref\gwdzki{K. Gawedzki, ``Topological Actions in Two-Dimensional
Quantum Field Theories,'' in {\it Nonperturbative
Quantum Field Theory}, G. 't Hooft, A. Jaffe, et. al. , eds. ,
Plenum 1988}
\lref\gmps{A. Gerasimov, A. Morozov, M. Olshanetskii,
 A. Marshakov, S. Shatashvili ,``
Wess-Zumino-Witten model as a theory of
free fields,'' Int. J. Mod. Phys. A5 (1990) 2495-2589
 }
\lref\gerasimov{A. Gerasimov, ``Localization in
GWZW and Verlinde formula,'' hepth/9305090}
\lref\ginzburg{V. Ginzburg, M. Kapranov, and E. Vasserot,
``Langlands Reciprocity for Algebraic Surfaces,'' q-alg/9502013}
\lref\giveon{hep-th/9502057,
 S-Duality in N=4 Yang-Mills Theories with General Gauge Groups,
 Luciano Girardello, Amit Giveon, Massimo Porrati, and Alberto
Zaffaroni
}

\lref\gottsh{L. Gottsche, Math. Ann. 286 (1990)193}
\lref\gothuy{L. G\"ottsche and D. Huybrechts,
``Hodge numbers of moduli spaces of stable
bundles on $K3$ surfaces,'' alg-geom/9408001}
\lref\GrHa{P.~ Griffiths and J.~ Harris, {\it Principles of
Algebraic
geometry},
p. 445, J.Wiley and Sons, 1978. }
\lref\ripoff{I. Grojnowski, ``Instantons and
affine algebras I: the Hilbert scheme and
vertex operators,'' alg-geom/9506020.}
\lref\adhmfk{I. Grojnowski,
A. Losev, G. Moore, N. Nekrasov, S. Shatashvili,
``ADHM and the Frenkel-Kac construction,'' in preparation}

\lref\hitchin{N. Hitchin, ``Polygons and gravitons,''
Math. Proc. Camb. Phil. Soc, (1979){\bf 85} 465}

\lref\hklr{Hitchin, Karlhede, Lindstrom, and Rocek,
``Hyperkahler metrics and supersymmetry,''
Commun. Math. Phys. {\bf 108}(1987)535}
\lref\hirz{F. Hirzebruch and T. Hofer, Math. Ann. 286 (1990)255}
\lref\hms{hep-th/9501022,
 Reducing $S$- duality to $T$- duality, J. A. Harvey, G. Moore and A.
Strominger}
\lref\johansen{A. Johansen, ``Infinite Conformal
Algebras in Supersymmetric Theories on
Four Manifolds,'' hep-th/9407109}
\lref\kronheimer{P. Kronheimer, ``The construction of ALE spaces as
hyper-kahler quotients,'' J. Diff. Geom. {\bf 28}1989)665}
\lref\kricm{P. Kronheimer, ``Embedded surfaces in
4-manifolds,'' Proc. Int. Cong. of
Math. (Kyoto 1990) ed. I. Satake, Tokyo, 1991}

\lref\KN{Kronheimer and Nakajima,  ``Yang-Mills instantons
on ALE gravitational instantons,''  Math. Ann.
{\bf 288}(1990)263}
\lref\krmw{P. Kronheimer and T. Mrowka,
``Gauge theories for embedded surfaces I,''
Topology {\bf 32} (1993) 773,
``Gauge theories for embedded surfaces II,''
preprint.}

\lref\hypvol{A. Losev, G. Moore, N. Nekrasov, S. Shatashvili,
``Localization for Hyperkahler Quotients,
Integration over Instanton Moduli,
and ALE Spaces,'' in preparation}
\lref\fdrcft{A. Losev, G. Moore, N. Nekrasov, S. Shatashvili, in
preparation.}
\lref\cenexts{A. Losev, G. Moore, N. Nekrasov, S. Shatashvili,
``Central Extensions of Gauge Groups Revisited,''
hep-th/9511185.}
\lref\maciocia{A. Maciocia, ``Metrics on the moduli
spaces of instantons over Euclidean 4-Space,''
Commun. Math. Phys. {\bf 135}(1991) , 467}
\lref\mickold{J. Mickelsson, CMP, 97 (1985) 361.}
\lref\mick{J. Mickelsson, ``Kac-Moody groups,
topology of the Dirac determinant bundle and
fermionization,'' Commun. Math. Phys., {\bf 110} (1987) 173.}

\lref\milnor{J. Milnor, ``A unique decomposition
theorem for 3-manifolds,'' Amer. Jour. Math, (1961) 1}
\lref\taming{G. Moore and N. Seiberg,
``Taming the conformal zoo,'' Phys. Lett.
{\bf 220 B} (1989) 422}
\lref\nair{V.P.Nair, ``K\"ahler-Chern-Simons Theory'', hep-th/9110042
}
\lref\ns{V.P. Nair and Jeremy Schiff,
``Kahler Chern Simons theory and symmetries of
antiselfdual equations'' Nucl.Phys.B371:329-352,1992;
``A Kahler Chern-Simons theory and quantization of the
moduli of antiselfdual instantons,''
Phys.Lett.B246:423-429,1990,
``Topological gauge theory and twistors,''
Phys.Lett.B233:343,1989}
\lref\nakajima{H. Nakajima, ``Homology of moduli
spaces of instantons on ALE Spaces. I'' J. Diff. Geom.
{\bf 40}(1990) 105; ``Instantons on ALE spaces,
quiver varieties, and Kac-Moody algebras,'' preprint,
``Gauge theory on resolutions of simple singularities
and affine Lie algebras,'' preprint.}
\lref\nakheis{H.Nakajima, ``Heisenberg algebra and Hilbert schemes of
points on
projective surfaces ,'' alg-geom/9507012}
\lref\ogvf{H. Ooguri and C. Vafa, ``Self-Duality
and $N=2$ String Magic,'' Mod.Phys.Lett. {\bf A5} (1990) 1389-1398;
``Geometry
of$N=2$ Strings,'' Nucl.Phys. {\bf B361}  (1991) 469-518.}
\lref\park{J.-S. Park, ``Holomorphic Yang-Mills theory on compact
Kahler
manifolds,'' hep-th/9305095; Nucl. Phys. {\bf B423} (1994) 559;
J.-S.~ Park, ``$N=2$ Topological Yang-Mills Theory on Compact
K\"ahler
Surfaces", Commun. Math, Phys. {\bf 163} (1994) 113;
S. Hyun and J.-S.~ Park, ``$N=2$ Topological Yang-Mills Theories and Donaldson
Polynomials", hep-th/9404009}
\lref\parki{S. Hyun and J.-S. Park,
``Holomorphic Yang-Mills Theory and Variation
of the Donaldson Invariants,'' hep-th/9503036}
\lref\pohl{Pohlmeyer, Commun.
Math. Phys. {\bf 72}(1980)37}
\lref\pwf{A.M. Polyakov and P.B. Wiegmann,
Phys. Lett. {\bf B131}(1983)121}
\lref\prseg{Pressley and Segal, Loop Groups}
\lref\rade{J. Rade, ``Singular Yang-Mills fields. Local
theory I. '' J. reine ang. Math. , {\bf 452}(1994)111; {\it ibid}
{\bf 456}(1994)197; ``Singular Yang-Mills
fields-global theory,'' Intl. J. of Math. {\bf 5}(1994)491.}
\lref\segal{G. Segal, The definition of CFT}
\lref\seiberg{hep-th/9407087,
Monopole Condensation, And Confinement In $N=2$ Supersymmetric
Yang-Mills
Theory, N. Seiberg and E. Witten;
hep-th/9408013,  Nathan Seiberg;
hep-th/9408099,
Monopoles, Duality and Chiral Symmetry Breaking in
N=2 Supersymmetric QCD, N. Seiberg and E. Witten;
hep-th/9408155,
Phases of N=1 supersymmetric gauge theories in four dimensions, K.
Intriligator
and N. Seiberg; hep-ph/9410203,
Proposal for a Simple Model of Dynamical SUSY Breaking, by K.
Intriligator, N.
Seiberg, and S. H. Shenker;
hep-th/9411149,
 Electric-Magnetic Duality in Supersymmetric Non-Abelian Gauge
Theories,
 N. Seiberg; hep-th/9503179 Duality, Monopoles, Dyons, Confinement
and Oblique
Confinement in Supersymmetric $SO(N_c)$ Gauge Theories,
K. Intriligator and N. Seiberg}
\lref\sen{A. Sen,
hep-th/9402032, Dyon-Monopole bound states, selfdual harmonic
forms on the multimonopole moduli space and $SL(2,Z)$
invariance,'' }
\lref\shatashi{S. Shatashvili,
Theor. and Math. Physics, 71, 1987, p. 366}
\lref\thooft{G. 't Hooft , ``A property of electric and
magnetic flux in nonabelian gauge theories,''
Nucl.Phys.B153:141,1979}
\lref\vafa{C. Vafa, ``Conformal theories and punctured
surfaces,'' Phys.Lett.199B:195,1987 }
\lref\VaWi{C.~ Vafa and E.~ Witten, ``A Strong Coupling Test of
$S$-Duality",
hep-th/9408074.}
\lref\vrlsq{E. Verlinde and H. Verlinde,
``Conformal Field Theory and Geometric Quantization,''
in {\it Strings'89},Proceedings
of the Trieste Spring School on Superstrings,
3-14 April 1989, M. Green, et. al. Eds. World
Scientific, 1990}
\lref\mwxllvrld{E. Verlinde, ``Global Aspects of
Electric-Magnetic Duality,'' hep-th/9506011}
\lref\wrdhd{R. Ward, Nucl. Phys. {\bf B236}(1984)381}
\lref\ward{Ward and Wells, {\it Twistor Geometry and
Field Theory}, CUP }
\lref\wittenwzw{E. Witten, ``Nonabelian bosonization in
two dimensions,'' Commun. Math. Phys. {\bf 92} (1984)455 }
\lref\grssmm{E. Witten, ``Quantum field theory,
grassmannians and algebraic curves,'' Commun.Math.Phys.113:529,1988}
\lref\wittjones{E. Witten, ``Quantum field theory and the Jones
polynomial,'' Commun.  Math. Phys., 121 (1989) 351. }
\lref\wittentft{E.~ Witten, ``Topological Quantum Field Theory",
Commun. Math. Phys. {\bf 117} (1988) 353.}
\lref\Witdgt{ E.~ Witten, ``On Quantum gauge theories in two
dimensions,''
Commun. Math. Phys. {\bf  141}  (1991) 153.}
\lref\Witfeb{E.~ Witten, ``Supersymmetric Yang-Mills Theory On A
Four-Manifold,'' J. Math. Phys. {\bf 35} (1994) 5101.}
\lref\Witr{E.~ Witten, ``Introduction to Cohomological Field
Theories",
Lectures at Workshop on Topological Methods in Physics, Trieste,
Italy,
Jun 11-25, 1990, Int. J. Mod. Phys. {\bf A6} (1991) 2775.}
\lref\wittabl{E. Witten,  ``On S-Duality in Abelian Gauge Theory,''
hep-th/9505186}
\lref\faddeevlmp{L. D. Faddeev, ``Some Comments on Many Dimensional Solitons'',
Lett. Math. Phys., 1 (1976) 289-293.}

\Title{ \vbox{\baselineskip12pt\hbox{hep-th/9509151}
\hbox{PUPT-1564}
\hbox{ITEP-TH.5/95}
\hbox{YCTP-P15/95}}}
{\vbox{
\centerline{Four-Dimensional Avatars }
\centerline{  of }
\centerline{ Two-Dimensional RCFT}}}\footnote{}
{Talk
presented at the USC conference
Strings'95, March 18, 1995 and at the Trieste
Conference on S-Duality and Mirror Symmetry,
June 8,1995}
\medskip
\centerline{Andrei Losev $^1$, Gregory Moore $^2$,
Nikita Nekrasov $^3$, and Samson Shatashvili $^{4}$\footnote{*}{On
leave of
absence from St. Petersburg Steklov Mathematical Institute, St.
Petersburg,
Russia.}}

\vskip 0.5cm
\centerline{$^{1,3}$ Institute of Theoretical and Experimental
Physics,
117259, Moscow, Russia}
\centerline{$^3$ Department of Physics,
Princeton University, Princeton NJ 08544}
\centerline{$^{1,2,4}$ Dept.\ of Physics, Yale University,
New Haven, CT  06520, Box 208120}
\vskip 0.1cm
\centerline{losev@waldzell.physics.yale.edu}
\centerline{moore@castalia.physics.yale.edu}
\centerline{nikita@puhep1.princeton.edu}
\centerline{samson@euler.physics.yale.edu}

\medskip
\noindent
We investigate a 4D  analog of 2D WZW
theory.
The theory turns out to have
surprising finiteness properties and
an infinite-dimensional current algebra symmetry.
Some correlation functions are determined by this
symmetry.
One way to define the theory systematically
proceeds by the quantization of moduli spaces of
holomorphic vector bundles over algebraic surfaces.
We outline how one can define vertex operators in
the theory. Finally, we define
four-dimensional ``conformal blocks'' and present
an analog of the Verlinde formula.

\Date{September 26, 1995;Revised December 19,1995}

\newsec{Introduction and Conclusion}

Two-dimensional rational conformal field theories
are completely solvable and have formed
the basis for much progress in understanding
quantum field theory. It  is therefore natural to
ask if there exist 4D QFT's which are as ``solvable''
as 2D rational conformal field theories. The aim of the present
paper is to outline just such a class of theories, we
will refer to them as 4D WZW theories, or
\wzwf, for short.
These theories have infinite dimensional symmetry
algebras generalizing those of the two-dimensional
WZW model.
Our results indicate that
a 4D analog of RCFT probably  exists.
Much
work remains to be done to flesh out this outline.

In constructing \wzwf\ one learns an important
philosophical lesson.  In cohomological
field theories \wittentft\Witr\CMR\
 a {\it topological
sector} is embedded in
a more complicated nontopological theory.
In the \wzwf\ theory close relations
to algebraic geometry rapidly become apparent,
and it appears that there is an intermediate
sector of the theory  which
is exactly solvable using
methods of algebraic geometry.
We refer to this sector as the
{\it algebraic sector} of the theory.
We postulate that there is a class of
{\it algebro-geometric quantum field
theories} analogous to
the class of topological field theories, in the sense mentioned
above.

In 2D RCFT the first order ``bc systems''
\ref\FMS{D. Friedan, E. Martinec,
and S. Shenker, ``Conformal Invariance,
Supersymmetry and String Theory,''
Nucl.Phys.B271:93,1986}
play an important role. An interesting
four-dimensional generalization of these
theories exists and, at least in the
algebraic sector, provides a four-dimensional
analogue of nonabelian bosonization
\fdrcft.

This note is a much-truncated version of a longer
manuscript \fdrcft, which contains fuller explanations,
more precise statements, and some results on
representation theory, fermionization, and symplectic
volumes not covered here.
One potentially important application of
the present work is to nonperturbative
string theory, since the theory we discuss
is related to the string field theory of the
$N=2$ string \ogvf\berk.

The results in
section 5.3  were obtained in
collaboration with Ian Grojnowski
and also appear in \ripoff.

\newsec{  $WZW_4$: Definition and  Properties of Classical Theory}

\subsec{Lagrangian.}

Let $X_4$ be a  four-manifold
equipped  with a
metric  $h_{\mu\nu}$ and a
closed 2-form $\omega \in \Omega^2(X;\IR)$.
Let $g\in Map(X_4, G)$ for a Lie group $G$.
Fix a reference field configuration $g_0(x)$.
Then, for any $g(x)$ in the same homotopy
class as $g_0(x)$ we may define
a natural analog of the D=2 WZW
theory by the Lagrangian:

\eqn\wzwiii{
S_\omega[g;g_0]=
{f_\pi^2\over  8 \pi} \int_{X_4}  {\Tr} g^{-1} d g \wedge * (g^{-1} d
g)
+ {i \over  12 \pi} \int_{X_5} \omega\wedge {\Tr}(g^{-1} d g)^3
}
Here  $f_\pi$ is a dimensionfull parameter.
\foot{Our conventions are the following: Differential
forms are considered dimensionless, but
$dx^\mu$ carries dimension one. Thus, the metric
$g_{\mu\nu}$ is dimensionless, but  $f_{\pi}^{2}$, $\omega_{\mu\nu}$
are
dimension
$-2$, etc.}
In the integral over
 $X_5= X_4 \times I $ in \wzwiii\
$\omega$ is independent of the fifth
coordinate; moreover, we use
a homotopy of $g$ to  $g_0$. The action is independent of the
choice of homotopy up to a multiple of the periods of $\omega$.
\foot{Note that $X_5$ is a cylinder, rather than a cone.
This is necessary since $X_4$ might not be cobordant
to zero, and since the periods of $\omega$ might
be nontrivial.  As was noted in \etingof, the
 latter fact caused difficulties in
finding a ``Mickelsson''-type construction \mick\  of
$\widehat {\rm Map}({\Sigma}, G)$ where $\Sigma$
is a Riemann surface. Using a cylinder and
a homotopy construction this problem can be
overcome  \cenexts.  A completely different solution
to this problem  has
recently been described in \fk.}

\subsec{Classical Equations of Motion. K\"ahler Point}

The classical equations of motion, following from \wzwiii, are:
\eqn\gcem{\eqalign{
& d * g^{-1} d g + \omega \wedge g^{-1} dg \wedge g^{-1} dg = \cr
& \p_{\mu}
h^{\mu\nu} \sqrt{h}
g^{-1} \p_{\nu} g  + \epsilon^{\alpha \beta \gamma \delta}
\omega_{\alpha \beta} g^{-1} \p_{\gamma} g g^{-1} \p_{\delta} g
=0\cr}}

They simplify
drastically in the case where
$X_4$ is a
complex four-manifold with
K\"ahler metric with $\omega$ the
associated K\"ahler form:
\eqn\kfrm{
\omega = {i  \over  2} f_\pi^2 h_{i \bar j} dz^i \wedge d z^\jb
}
We refer to this point in the space of Lagrangians
as the ``K\"ahler point.''

Using standard properties of the Hodge
star
we may rewrite the action as:
\eqn\wzwiiv{
S_\omega[g]=
-{i \over  4 \pi} \int_{X_4}
\omega\wedge  {\Tr} \bigl(g^{-1} \p g \wedge g^{-1} \pb g\bigr)
+ {i \over  12 \pi} \int_{X_5} \omega\wedge {\Tr}(g^{-1} d g)^3
}

The equations of motion following from
\wzwiiv\ are:
\eqn\wzwv{
\eqalign{
\omega\wedge\pb\bigl(\gpg \bigr) & = 0 \cr}
}
These equations are known as the Yang equations,
and are equivalent to the self-dual Yang-Mills
equations.

{\bf Remarks .}

\item{1.} The Lagrangian \wzwiiv\  was
first written by Donaldson  \dnld.
It was studied by Nair and Schiff
as a natural generalization
of the 2D CFT/ 3D CSW correspondence \ns\nair.

\item{2.} In \ogvf \berk\  the special K\"ahler point is related to
 the  classical field theory of the $N=2$ string.
String  investigations of this theory
have focused on the $S$-matrix for $\pi$ defined
by $g=e^\pi$. The present paper studies
field configurations $g$ related to instantons.
Hence one may expect that results of this
 paper will have a bearing on nonperturbative
$N=2$ string theory.

\subsec{Twistor transform and Classical Integrability}

We explain how one can solve equations
of motion in $WZW_{4}$ theory
 using the twistor transform of the
self-dual Yang-Mills equations.

$\IR^4$ can be endowed with complex structures
parametrized by $\IP^1$. Choosing a basepoint
complex structure $z^1,z^2$ the others
are defined by:
$ z^A_u= z^A+ u \epsilon^{AB} \zb^B
$
and
$u$ labels a point  $ u\in \hat\IC\cong \IP^1 \cong SO(4)/U(2)$.

Given a region $\CR\subset \IR^4$
its twistor space $\hat \CR\to \CR$
has as fiber the sphere $\IP^1$ of complex structures
compatible with an orientation.
The twistor transform defines a correspondence
between solutions of the Yang equation $\CY(\CR)$ and
twistor data $\CT(\hat \CR)$ defined by:
\eqn\twstrdfs{
\eqalign{
\CY(\CR) &  \equiv \{ g: \omega\wedge\pb\bigl(\gpg \bigr)  = 0 \quad
{\rm on} \quad \CR\} \cr
\CT(\hat \CR)  &  \equiv
\{ G(s^1,s^2, u): \hat \CR_+\cap \hat \CR_-\to GL(n,\IC)\}
\cr}
}
where  $\hat \CR_\pm$ are patches defined by
the north/south pole and:

\item{1.} $G(z^1_u, z^2_u,u) = H_-^{-1}(x,u) H^+(x,u)$ for
$0< \vert u \vert < \infty$.

\item{2.}  $H_\pm(x,u)$  holomorphic in $u$ for $\mid u\mid <\infty,
\mid u\mid >0$

Briefly, choosing a gauge
$
A^{(1,0)}=0 , A^{(0,1)}=-\pb g g^{-1}
$
the SDYM are satisfied iff $\forall u,  F^{(0,2),u} =0 $
which  holds iff $\pb^{(0,1),u}_{\bar A} H(x,u)=0$.
Imposing holomorphy of  $H(x,u)$ in $u$ forces us to
choose two functions $H_\pm$ holomorphic
on the patches $\hat \CR_\pm$ and related
as in item 1 above. We then identify
$g  = H_+(x,u=0)$ as a solution to the Yang equation (this
construction
is analagous to the one in \blzh).

Note that all solutions to the Yang equation could be
obtained by taking different holomorphic functions $G$ and
solving the Riemann-Hilbert problem (1.) In this sense
the classical
equations of motion for a complex group $G$ are integrable.

It is also worth noting that the twistor transform
gives a generalization to 4D of the important
property of holomorphic factorization in
\wzwt\ \fdrcft.
\subsec{Canonical Approach to the Classical Theory and Current
Algebra}
We consider a four-manifold with
space-time splitting
$X_4= X_3 \times \IR$.
\foot{We continue to work with
Euclidean signature}
The phase space of the model can
be identified with  $\CP= T^* Map(X_3,G)$,
where the momenta are valued in:
 $I^a(x) \in \Omega^3(X_3,\lieg)$
and $\lieg$ is the Lie algebra of $G$.

Writing the action  $S_\omega$ in first order form
we extract the symplectic form:
\eqn\smpfrm{
\Omega_\omega = \int_{X_3} {\Tr}\Biggl[
\delta I \wedge g^{-1} \delta g - \biggl(I + {1 \over  4 \pi}
\omega\wedge
g^{-1} d g\biggr) (g^{-1} \delta g)^2 \Biggr]
}
from which we obtain the commutation relations of
$I^a(x), g(x)$.
\eqn\cncmrl{
\eqalign{
[I^a(x), I^b(y)]_\omega & = f^{ab}_c (I + {1 \over  4 \pi}
\omega\wedge
g^{-1} d g )^c \delta^{(3)}(x-y) \cr
[I^a(x), g(y) ]_\omega & = g(y) T^a \delta^{(3)}(x-y) \qquad . \cr}
}
{}From these relations we can obtain the generalization
of two-dimensional current algebra.
Form the combination
\eqn\curr{
J^a(x) = I^a(x) - {1 \over  4 \pi} {\omega} \wedge
(g^{-1} d g) ^a (x)
}
For a  $\lieg$-valued function on $X_3$
$\epsilon^a(x)$ it gives rise to the charge:
\eqn\chrg{
Q(\epsilon) = \int_{X_{3}} {\Tr} (\epsilon J)}
The charges $Q(\epsilon)$ obey the following
commutation relations:
\eqn\pbr{
	\{ Q({\epsilon}_{1}),Q( {\epsilon}_{2}) \} =
	Q( [ {\epsilon}_{1}, {\epsilon}_{2}])
        + \int_{X_{3}} \omega \wedge {\Tr} (\epsilon_{1} d
{\epsilon}_{2} )
}
We denote this algebra as
$\kappa(X_3,\lieg,\omega)$.

{\bf Remarks}.

\item{1.} There are several possible
generalizations of 2D current algebra.
The above algebra is the one  relevant to
the {\it algebraic} sector
of our theory, but,
for example,
using the commutation relations \cncmrl\ it is
possible to form a larger algebra as follows.
We can form the objects like
$I_{\xi} = I - {1 \over  4 \pi} ({\omega} + {\xi}) \wedge
g^{-1} d g$, where $\xi$ stands for {\it any} two-form on $X_{3}$.
These form
an algebra of charges $Q_{\xi}(f) = \int_{X_{3}} {\Tr} (f I_{\xi})$:
\eqn\bigal{
\eqalign{
\{ Q_{\xi_{1}}(f_{1}), Q_{{\xi}_{2}} & (f_{2}) \}  = Q_{ {\xi}_{1} +
{\xi}_{2}}([f_{1}, f_{2}])  +
\cr
&+  \int_{X_{3}}
\Biggl[ ( {\omega} + {\xi}_{1} ) {\Tr} ( f_{1} df_{2}) -
 ( {\omega} + {\xi}_{2} ) {\Tr} ( f_{2} df_{1})
\Biggr]\cr}
}

\item{2.} There is an interesting similarity between
the algebra $\kappa(X_3,\lieg,\omega)$
and the  algebra discovered
in  \fs\fsi\mickold\fz\
for the case when an abelian gauge field  strength  is equal to a
K\"ahler form $\omega$. This analogy suggests the existence
of a free field
interpretation of the algebraic sector (see below) of
\wzwf. Indeed, such an interpretation
of the algebraic subsector of the theory
exists and will be described in \fdrcft.

\newsec{$WZW_{4}$: Quantum theory}

\subsec{Quantization of $[\omega]$
and the first sign of algebraic geometry}

As usual, the coefficient of the WZ term is quantized.
Two different homotopies of $g$ to
$g_0$  define a map
$ g: S^1\times X \to G$.
Consequently, if the group $G$ is
nonabelian,
 the measure $\exp i S$ in the
path integral is only
well-defined if
\eqn\contii{
\omega\wedge
{1 \over  12 \pi} {\Tr}(g^{-1} d g)^3\in H^5(S^1\times X_4;2\pi \IZ)
}
which forces the cohomology class
 $[\omega]$ to lie in the lattice:
\eqn\contiii{
[\omega] \in H^2(X_4; \IZ)
}
The class $[\omega]$ is the
four-dimensional analog of $k$.
Note that although $[\omega]$ is quantized
the Lagrangian depends on the representative
of the class.
Since $\omega$ is of type $(1,1)$
condition \contiii\  implies
$[ {\omega}] \in H^2(X;\IZ)\cap H^{1,1}(X;\IR)$
so the metric is Hodge, and,
by the Kodaira embedding theorem,
if $X_4$ is compact, it
must be  algebraic \GrHa.
\foot{Curiously, the supersymmetric
$\sigma$ model can only be coupled
to $N=1$ supergravity when the target
is a  Hodge manifold \bagger, for similar
reasons.}

\subsec{Remarks on States and Hilbert Spaces:
An operator formalism in 4d}

One goal of this investigation is the generalization of
concepts of 2D CFT to 4D. In the 2D case one
fruitful approach to defining states and Hilbert
spaces proceeds by considering states defined by
path integrals on 2-folds with boundary, the
boundary is interpreted  as a spatial  slice.

In 2D  a compact connected
spatial slice is necessarily a copy of
$S^1$.
In 4D,
on the other hand,
there is a wide variety of possible notions of
space.  Two natural choices are $S^3$, corresponding
to radial propagation around a point and
a circle bundle over an embedded surface,
corresponding to radial propagation in
the normal bundle to the surface.

Suppose now that
the 3-fold bounds a 4-fold $Y = \p X$, then
a particular state $\Psi_Y\in \CH(Y)$ is
defined by the path integral. An important
problem is to describe how one can construct
this state explicitly. One immediately
encounters a fundamental difficulty in
defining the state via conserved charges
called the ``2/3 problem.''

Conserved charges in 2D come from (anti-) holomorphic
currents. If $f\vert_{X_1}$ extends to
a holomorphic function on $X_2$ then by contour
deformation:
\eqn\contdef{
\pb(\gpg)=0 \quad \Rightarrow\quad
\oint_{X_1} {\Tr}(f \gpg) \Psi =0
= \oint_{X_1} {\Tr} (\bar f \pb g g^{-1}) \Psi
}
In 2D, ``enough'' boundary values $f$ extend to
holomorphic or anti-holomorphic functions on $X_2$
that the conserved charges determine the state up to
a finite-dimensional vector space (related to
conformal blocks).

In 4D, $\omega\wedge \gpg$ is a $\pb$-closed
$(2,1)$ form, and hence,  if
$f\vert_{X_3}$ extends to
a holomorphic function on $X_4$  ``contour deformation''
of $X_3$ implies
\eqn\contdefi{
\oint_{X_3} {\Tr}\biggl[ f(z^1,z^2) \omega\wedge
\gpg\biggr] \Psi =0
}
but note there are not nearly enough such functions
to determine the vacuum. Even in the best cases
\foot{for example, in radial quantization around
a point}
 we are missing one {\it functional} degree of freedom:
boundary values of holomorphic functions depend on
two functional degrees of freedom but to determine
the state we need conserved charges depending on
three functional degrees of freedom.

One possible approach to the problem uses the twistor transform
of the SDYM equations. In the abelian case it does
solve the problem of missing charges - harmonic functions on
a ball with the flat
metric are linear combinations of holomorphic functions in
different complex structures.

\subsec{The perturbative vacuum is one-loop finite}

The \wzwf\ theory possesses unusual
finiteness properties. The one-loop renormalization of the vacuum
state
may be carried out via the background-field
method. We let  $g=e^{i \pi/f_\pi} g_{cl}$ where
$g_{cl}$ is a solution of the classical equations
of motion with given boundary conditions.   For
one-loop renormalization it suffices to keep
the terms quadratic in $\pi(x)$:
\eqn\piexp{
S= S_{cl} + \int  (\nabla_\mu(J_{cl}) \pi)^2  + \pi^a  M^{ab} \pi^b
+ \CO(\pi^3)
}
where the connection $\nabla_\mu(J_{cl})$ and
$M^{ab}$  are constructed
from the classical solution of the equation of motion.
The divergent terms at one-loop may be
extracted from the Seeley expansion:
\eqn\seeley{
\Delta S=\int {1 \over  \epsilon^4} C +
 {1 \over  \epsilon^2} {\Tr} M + (\log \epsilon) {\Tr} \Biggl[
{\half} M^2
+ {1 \over  6} \bigl({\nabla}^2 M  -{\half} F_{\mu\nu}(J)^{2} \bigr)
\Biggr]
+\cdots
}
where $C$ is a $\pi$- independent constant
and $F_{\mu\nu}(J)$ is a fieldstrength constructed from
$\nabla_\mu(J_{cl})$.
For the $WZW_4$ action we find
that
$$
M=0, \quad {\bar \nabla}(J_{cl})={\bar \partial},\quad
{\nabla}(J_{cl})={\partial}+[g^{-1}_{cl}{\partial}g_{cl}, \cdot ].
$$
We conclude that there is no
quadratic divergence at the K\"ahler point
and in addition,
using the equations of motion
we see that  the
logarithmic divergences also vanish
once boundary terms are properly included.
Of course, these statements can also
 be checked directly using Feynman diagrams.

\subsec{Discussion of Renormalizability}
Given the surprising one-loop finiteness one
may naturally wonder if the model is
finite.  This remains a mystery.
Several arguments
indicate that the renormalizability
properties of the theory are special, but
eventually they remain inconclusive.

Even if the theory is 2-loop infinite we should not
give up.
\wzwf\ is clearly a very distinguished
theory. In particular, as we shall see,
the relation to the moduli of
instantons is completely parallel to that of \wzwt\
to the moduli space of flat connections, and one can
elevate this statement to a defining principle of the
quantum theory. Thus, even if it cannot be defined by standard
field-theoretic techniques, we believe the theory
exists. This, in fact, is one of the sources of
its great interest. Therefore, in the sequel, we choose another way
of
defining the correlation functions.

\subsec{The PW formula and the quantum equations of motion}

At the ``K\"ahler point''
 the classical equations of motion have a local ``two-loop group''
symmetry
$\CH G_\IC \times \CH G_\IC$,
where $\CH G_\IC = \{ g(z^1,z^2)\in G_\IC \} $,
which takes
$g\to g_L(\zb^i) g(z,\zb) g_R(z^i)$
As in 2D this  is related to the
Polyakov-Wiegman (PW) formula \pwf:
\eqn\genpwi{
S_\omega[gh]= S_\omega[g]  + S_\omega[h]
-{i\over  2 \pi} \int_{X_4} \omega\wedge
{\Tr} [\gpg \pb h h^{-1}]
}
Assuming invariance of the path integral measure $Dg$
with respect to the left action
of the  group ${\rm Map}(X,G)$ (with $G$ compact)
we can derive the
quantum equations of motion:
\eqn\qem{\eqalign{
& \pb \langle J \rangle = 0\cr
& J = \omega \wedge g^{-1} \p g\cr}}
\subsec{Ward identities}

The identity \genpwi\ implies that
the $(2,1)$-form current
$J=\omega\wedge \gpg$ satisfies:
\eqn\wrdidsi{
\eqalign{
{i \over  2 \pi} \pb_x\biggl
\langle J^A(x) \prod J^{A_i}(x_i)\biggr\rangle &
= \sum_i (\omega\wedge \p)_i  (\delta(x,x_i)) \delta^{A, A_i}
\biggl \langle  \prod_{j \not= i}  J^{A_j}(x_j) \biggr\rangle \cr
+ \sum_i f^{A A_i B_i}  &  (\delta(x,x_i))
\biggl \langle  J^{B_i }(x_i)
 \prod_{j \not= i}  J^{A_j}(x_j)\biggr\rangle \cr}
}
where $\delta(x,y)$ is a $4$-form in $x$ and
a $0$-form in $y$.

In strong contrast to the situation in 2D, the
identities \wrdidsi\ do {\it not} determine
the correlation functions since  $\pb$ has an
$\infty$-dimensional kernel and these identities
cannot fully determine the correlation functions.
However, the identities, together with simple
analyticity arguments do determine many correlation
functions in a way analogous to the 2D case.
 Indeed,
suppose $Y$ is a compact three-manifold, which
divides $X_{4}$ in two parts: $X_{+}$ and $X_{-}$,
and suppose that  $f$ is a function on $Y$ taking
values in the Lie algebra $\lieg$ and
 extending holomorphically
 to $X_{+}$. Then
 the integral
\eqn\vfy{V(f,Y) = \int_{Y}  {\Tr} (f J),}
 generates  an infinitesimal gauge transformation. The field $g$ in
the region
$X_{+}$ gets transformed to $g + fg$, and
remains unchanged in $X_{-}$. Now, keeping in mind
the fact, that $J$ transforms as
\eqn\jtas{J \to  J + \omega \wedge \p f + [ J, f],}
we can derive the correlation functions of the currents, integrated
with
appropriate functions $f$. Examples will be presented in the next
section.

\subsec{Definition of ${\CZ}({\bar A})$ and its equivariance
properties}
It is natural to consider the generating function
for the correlators of the currents $J$:
\eqn\zba{
{\CZ}({\bar A}) = \int Dg e^{- S_{\omega}(g) } e^{- {i \over  2 \pi}
\int_{X_{4}} {\Tr} J {\bar A}}
}
The generating parameter $\bar A$ is a
$\lieg\otimes \IC$-valued
$(0,1)$ form on $X_{4}$.
\foot{$\bar A$ is actually a connection, but we will work
on a topologically trivial bundle until section 4.4 below.}

Leaving aside the question of the regularization of the
quantity on the right hand side of \zba\  we
observe the following equivariance property of $\CZ$, following
from the PW formula:
\eqn\eqvrc{
\CZ (\bar A^{h^{-1}})
= e^{S_\omega(h) + {i\over{4{\pi}}} \int_{X_{4}} \omega
\wedge {\Tr} h^{-1} \p h {\bar A}} \CZ (\bar A)
}

Taking $h$ to be
infinitesimally close to $1$ we get the following
functional equation on $\CZ$:
\eqn\foo{
{\widehat F}^{(1,1)+}{\CZ} =
\omega \wedge F( \Lambda {{\delta}\over{\delta {\bar A}}}, {\bar A})
{\CZ}
= 0}
where
$$
F(A, {\bar A}) = \p {\bar A} - \pb A + [ A, {\bar A}]
$$
and $\Lambda$ is an operation inverse to multiplication by $\omega$.
Namely, the action of $\Lambda$ on a three form
 $\Omega_{\mu \nu \lambda}$
gives a one form
$$
(\Lambda \Omega)_{\mu}= {\half} \omega^{\nu \lambda}
\Omega_{\mu \nu \lambda},
$$
 where the bivector $\omega^{\nu \lambda}$ is inverse to
 $\omega_{\mu \lambda}$.
Moreover,  since
$J$ satisfies the flatness condition:
\eqn\flti{
\p (\Lambda J) + (\Lambda J)^{2} = 0,
}
$\CZ$ obeys
\eqn\ftz{
{\widehat F}^{2,0} {\CZ} = F^{2,0}
({\Lambda}{{\delta}\over{{\delta} {\bar A}}})
{\CZ} = 0
}

\newsec{Algebraic Sector}

In this section we assume that the quantum
theory can be defined preserving
invariance of the measure $Dg$.

\subsec{Definition of the algebraic sector }

{}From the PW identity we get
\eqn\wis{
\Biggl\langle \exp\biggl[  -{i\over  2 \pi}
\int_{X_4} \omega\wedge
{\Tr} [\gpg \pb h h^{-1}\biggr] \Biggr\rangle
= \exp\biggl[S_\omega[h] \biggr]
}
This identity is closely related to the algebraic
geometry of holomorphic vector bundles on
$X_4$ and characterizes the content of the
algebraic sector of the theory.
In the following subsections we will extract some
interesting information from this general statement.

Recall that ${\CZ}({\bar A})$ satisfies the
two functional equations \foo\ and \ftz.
Unfortunately,  in the
non-abelian theory the equation ${\widehat F}^{2,0}{\CZ}=0$
is non-linear
and is apparently insoluble. Therefore, we shall
restrict
${\CZ}$  to the space of connections  $\bar A$
such that $F^{0,2}({\bar A}) = 0$.  (Such connections
will be called   K\"ahler gauge fields and
the space of such connections is denoted
as ${\CA}^{1,1}$.)

{\it Our definition of the algebraic sector of $WZW_{4}$  theory is
as follows:
it is the set of correlators, which
can be extracted from the properties of ${\CZ}({\bar A})$, evaluated
on
${\CA}^{1,1}$}.

Naively,
one can argue that
the solution of the equation ${\widehat F}^{2,0}{\CZ}=0$ is
uniquely determined by the restriction of $\CZ$
onto the subspace ${\CA}^{1,1}$, and thus all current
correlators are determined by the algebraic sector.

\subsec{Simplest examples of algebraic correlators:
``Divisor current''  correlators}
In $WZW_2$ we study correlators of currents at points $z_1, \ldots,
z_n$, i.e.
$$
\langle j(z_1) \ldots j(z_n) \rangle.
$$
Note that points are divisors in
one dimensional complex geometry. For example, on ${\IC}{\IP}^{1}$
a  point
$z_1$ is the polar divisor of the function
$
f_1(z)=1/(z-z_1),
$
and thus,
\eqn\cchy{
j(z_1)=\int_{Y_1} j(z)f_1(z).
}
where $Y_1$ is a small circle around the point $z_1$.
As in \vfy\jtas\ this observable generates gauge
transformations and hence its correlators are
easily computed on ${\IC}{\IP}^{1}$.
 Equivalently, the generating function of current correlators is
 a partition function in the presence of a  field $\bar{A}$.
This field  defines the
structure of a holomorphic vector bundle on an algebraic curve.
 If  the bundle is holomorphically
trivial  we may find the correlators from the PW
formula. Even if the bundle is nontrivial, but
the curve is ${\IC}{\IP}^{1}$ we may use
singular gauge transformations (discussed below)
to find the correlators.
If the genus of the curve
 is greater than zero, the PW formula reduces the
problem of the computation of
 current correlators to the problem of the quantization of
the moduli space of holomorphic vector bundles.

 In complex dimension two,
arbitrary  correlators of currents  are generated
 by the function $\CZ(\bar{A})$ for unrestricted $\bar{A}$, and hence
cannot be computed within
 the algebraic sector. Nevertheless,
 it is possible to use the current to define an
observable at a divisor
 $D$ by analogy with the previous case, i.e.
 consider a meromorphic function $f\in \lieg\otimes \IC$
on $X_4$ such that its polar
divisor is $D$. Consider the three-dimensional boundary $Y$
of a small  tubular neighborhood  around the divisor.
Let us define an observable
 $J(f, D)$, associated to $(f,D)$ as
\eqn\divop{
 J(f, D)=\int_{Y} {\Tr}( f J )
}
This is just \vfy\ above, and only depends on $f, D$.

Correlators of the observables \divop\
 are generated by connections $\bar{A}\in \CA^{1,1}$,
and are therefore in the algebraic sector.
Moreover,
correlators of the observables \divop\ can be
computed as follows.  Suppose that the divisors
$D_i$ do not intersect one another so that
their tubular neighborhoods do not intersect.
Each observable generates holomorphic
gauge transformations outside its tubular
neighborhood  and since the current transforms
like a connection the  ``2-divisor" correlator is:
\eqn\twptfn{
\biggl\langle J(f_1, D_1) J(f_2, D_2) \biggr \rangle
=
\int_{Y_2} {\Tr} \biggl( f_2 \omega\wedge \p f_1\biggr)
}
Similarly, the "3-divisor" correlator is:
\eqn\thrptfn{
\eqalign{
\biggl\langle J(f_1, D_1) J(f_2, D_2) J(f_3, D_3)\biggr \rangle
 & =
 \int_{Y_2}
{\Tr} \Biggl\{ f_3 \omega\wedge \p \biggl( [f_1,f_2]
\biggr)\Biggr\} \cr
 &\qquad  +
 \int_{Y_3}
{\Tr} \Biggl\{ f_2 \omega\wedge \p \biggl( [f_3,f_1]
\biggr)\Biggr\} \cr}
}
The formulae \twptfn\thrptfn\  generalize
 current algebra on algebraic curves to current
algebra on algebraic surfaces.
Similarly the $n$-divisor correlators can be
written as above.

{\bf Remark}. One could also define observables
analogous to \divop\
associated to arbitrary divisors, rather than
polar divisors of  meromorphic functions.
These are still in the algebraic sector but their
correlators cannot be computed as easily since
they require information about nontrivial conformal
blocks.
\foot{Compare with the current correlator
on a Riemann surface of genus $>0$ in \wzwt. }

\subsec{Algebraic Correlators in the Abelian Theory}

In some two-dimensional conformal field theories one
can compute not only current correlators but also
vertex operator correlators using singular gauge
transformations.
\foot{See, for example
\bost\agmv\grssmm. In \grssmm\ these
relations were called
``multiplicative Ward identities.''}
In the 4D $U(1)$ \wzwf\ theory (which is
just a Gaussian model) this method may be
generalized as follows.
\foot{These remarks are closely related to
ref.  \banks. Another related example is given by
Maxwell theory on $X_4$. Singular gauge
transformations shift the line bundle and allow
one to produce a classical partition function
analogous to the partition functions of
2d gaussian models. They also allow one to
study the vertex operators
$\CV(D) = \exp \biggl[ {i \over  2 \pi} \int_D F \biggr] $
and
$\CV^*(D) = \exp \biggl[ {i \over  2 \pi} \int_D *F \biggr] $.
This and related issues have  been independently studied in
\wittabl\mwxllvrld. }

We would like to make a singular complexified
gauge transformation by a meromorphic function
$f$:
\eqn\scpgt{
g=e^{i \phi} \rightarrow f g f^\dagger
}
Intuitively, along the zero and
polar divisors of $f$ the effect of the singular gauge
transformation is to insert a vertex operator.
We can make this idea precise and compute the
correlators using the (left- and right-)  PW formula
by using a regularized version of $f$.
Suppose $div(f) = \sum n_i D_i$, and, near $D_i$,
we may describe the divisor in local coordinates
as $z_i =0$. Thus
we have
\eqn\neard{
f= z_i ^{n_i}(f_i + \CO(z_i) )
}
where $f_i$ can be meromorphic on $D_i$.
We now use the K\"ahler metric to choose tubular
neighborhoods $T_i$ around $D_i$ of radius
$\epsilon_i$, and use a partition of unity to
define a regularized gauge transformation with
parameter $f_\epsilon$  such that
$\vert f_\epsilon\vert^2  = \vert f\vert^2$
outside $\cup T_i$,  while
$\vert f_\epsilon\vert^2=(\vert z_i\vert^2 + \epsilon_i^2)^{n_i}
\vert f_i + \cdots \vert^2$ in $T_i - \cup_{j\not= i} T_j$ etc.

Now consider the effect of such a gauge transformation,
in the left-right version of the  PW formula \wis. The
LHS of  this  formula  involves a change of
action by:
\eqn\sngtr{
S_\omega \rightarrow S_\omega +
\sum_{i}   {i} n_i \int_{D^i} \omega \phi +\CO(\epsilon)
}
as $\epsilon\rightarrow 0$. So the LHS
formally becomes a correlator of operators
\eqn\abvos{
V_k[D] = e^{ {i}  k  \int_{D}\omega \phi}
}
Computing the RHS of this formula we find, as
$\epsilon\rightarrow 0$
\eqn\sngti{
S_\omega[\log \vert f_\epsilon\vert^2] \rightarrow {1 \over  2}
\sum_i \bigl[
n_i^2\log\epsilon_i^2  \int_{D_i} \omega  + n_i
\int_{D_i} \omega \log\vert f_i\vert^2 \bigr]
}
Exactly as in two dimensions the logarithmically
divergent terms are  renormalization
factors needed to normal-order the observables
\abvos. These observables accordingly have
anomalous dimension
\eqn\anmdm{
\Delta_k =
{k^2 \over  2}  \int_{D} \omega
}
and the renormalized operators have the
correlation function:
\eqn\ablans{
\biggl\langle \prod V_{n_i}[D_i] \biggr \rangle
=
\exp\biggl[\sum_i {n_i \over  2}
\int_{D_i} \omega \log\vert f_i\vert^2 \biggr]
}

{\bf Remark}.
We expect that the correlator will factorize
holomorphically, as in two-dimensions,
and that it should be possible to define
``chiral vertex operators.'' One may
expect a relation like:
\eqn\lftph{
\exp[\int_{D_0} ord_{D_0}(f) \omega \phi_{L}]
\exp[\int_{D_\infty} ord_{D_\infty}(f) \omega \phi_{L}] \sim
\exp \int_{T} \omega \wedge \p \phi
}
where the zero and polar divisors form
the boundary of a
three-manifold $T$:
$D_0-D_\infty = \p T$.  The details of
this proposal, especially in the nonabelian case,
appear to be nontrivial and
have not been worked out yet.

\subsec{Moduli of Vector Bundles and ASD Connections}

In two dimensions one of the possible ways of defining operators in
the theory
is through
coupling the basic field $g$ to non-trivial gauge fields.
For example, in the abelian theory,
by coupling to a line bundle $\CL$
with non-trivial $c_{1}$ one can get insertions
of operators at the  points corresponding to the divisor of $\CL$.
This procedure can be generalized to the
nonabelian case. Therefore, it is natural
to try to generalize the $WZW_{4}$ action for the theory in the
background of a
non-trivial gauge field.

Let $E\rightarrow X_4$ be a
rank $r$ complex hermitian
vector bundle with metric $(\cdot,\cdot)_E$ and
Chern classes $c_1,c_2$ and let
$g(x)$ be
a section of $Aut({E})$.
To write a well-defined action
generalizing \wzwiiv\ we must introduce a
connection $\nabla$  on $E$. It gives rise to a connection
on $Aut({E})$, which we also denote as $\nabla$.
Since $X_4$ is complex
we can split the connection into its $(1,0)$ and
$(0,1)$ pieces:
${\nabla} = {\nabla}^{(1,0)} + {\nabla}^{(0,1)}$.
The  $WZW_{4}$ action in the background
$(E, {\nabla})$ is (we use here the result of \faddeevlmp):
\eqn\ggdact{
\eqalign{
S_{{\omega}; E, {\nabla}} [g]  =
-{i \over  4 \pi} \int_{X_4}
\omega\wedge  {\Tr} \bigl(g^{-1} {\nabla}^{(1,0)} g \wedge g^{-1}
{\nabla}^{(0,1)} g\bigr)+
\cr
+ {i \over  12 \pi} \int_{X_5} \omega\wedge \biggl[ {\Tr }
(\tilde g^{-1} \tilde{\nabla} \tilde g)^3
+ 3 {\Tr } \ F_{\tilde{\nabla} }[\tilde g^{-1}  \tilde \nabla \tilde g
+
(\tilde \nabla \tilde g)  \tilde g^{-1}  ] \biggr]
\Biggr)\quad .}
}
In the formula we trivially extended $E$ to $X_{5}$ and we denote the extension
of $g$, $\nabla$ to $X_{5}$ as $\tilde g$, $\tilde \nabla$.

In order for the Polyakov-Wiegmann formula to be generalizable
to non-trivial $E$ we must check whether
$S_{{\omega}; E, {\nabla}}[gh] - S_{{\omega}; E, {\nabla}}[g]-
S_{{\omega}; E, {\nabla}}[h]$
is local in four dimensions. After a simple computation
one obtains:
\eqn\pwe{\eqalign{
S_{{\omega}; E, {\nabla}}[gh] = S_{{\omega}; E, {\nabla}}[g]
+
S_{{\omega}; E, {\nabla}}[h]  \cr
-{i\over{2\pi}}
\int_{X_{4}} \omega \wedge {\Tr} \bigl[ g^{-1} ({\nabla} ^{(1,0)}g )
({\nabla}^{(0,1)}h) h^{-1}\bigr] }
}
As before we define
\eqn\oneone{
\CA^{1,1}(E\to X_4) \equiv \{ {\nabla}:
F^{0,2}=F^{2,0} =0\}
}
to be the space of unitary connections whose curvature is
of type $(1,1)$. If ${\nabla} \in \CA^{1,1}$ then
$\nabla^{(0,1)}$ is the Dolbeault operator defining an
integrable holomorphic structure $\CE$
on the vector bundle $E$. The moduli space of
holomorphic vector bundles
\foot{Technically one should specify carefully
a compactification of this moduli space.
One reasonable possibility is to choose
 the moduli space
of rank $r$
torsion free sheaves on $X$ semistable
with respect to a polarization
$H$.}
is given by:
$\CH\CB \equiv  \{ {\nabla}^{(0,1)} = {\nabla}_{0}^{(0,1)} + \bar A:
({\nabla}^{(0,1)})^2=0 \}/Aut({E}) $, where ${\nabla}_{0}^{(0,1)}$ is
some reference connection
and $\bar A$ is a $(0,1)$-form with values in $ad({E})$.

We will perform a functional  integral over  the group of unitary
automorphisms $Aut(E)^{u}$
of the bundle $E$ .
Once again we consider  the generator of
current correlation functions. For $\bar A \in
\Omega^{0,1}(ad({E}))$:
\eqn\holvbi{
{\CZ}_{E} [\bar A] = \biggl\langle \exp {i \over  2 \pi}
\int_{X_4} \omega\wedge
{\Tr} \bar A  g^{-1} {\nabla}_{0}^{(1,0)} g  \biggr\rangle
}
By the generalized PW formula \pwe\ we have the equivariance
condition:
\eqn\eqvrce{
{\CZ}_{E}(\bar A^{h^{-1}}) =
e^{S_{\omega; E, {\nabla_{0}}} [h] + {i\over{2\pi}}
\int_{X_{4}} {\omega} \wedge {\Tr} {\bar A}  h^{-1} {\nabla}_{0}^{(1,0)} h
}
{\CZ}_{E}(\bar A)
}
where ${\bar A}^{h^{-1}} = h {\bar A} h^{-1} - (  {\nabla}_{0}^{(0,1)}h  )
h^{-1}$.
Restricting  $\nabla_{0}^{(0,1)} + {\bar A}$ to be in
$\CA^{1,1}(E\to X_4)$  we learn that $\CZ_E [\bar A] \in H^0(\CH\CB;
\CL_\omega)$
where the line bundle $\CL_\omega$ has non-trivial first
Chern class
$c_1(\CL_\omega)  = \omega_\CM$, associated with the
K\"ahler form
on ${\CA}^{1,1} ( E \rightarrow X_{4})$;
$ \omega_\CM= \int_X \omega_X \wedge  {\Tr} \delta A \wedge
\delta A $. In other words, $\CZ_E [\bar A] $ is a
wavefunction for the quantization of $\CH\CB$.

The quantization can also be understood as
the quantization of the moduli space
$\CM^+$ of ASD connections.
Mathematically, this follows from the
Donaldson-Uhlenbeck-Yau theorem
\foot{which plays the role  in four dimensions of
the Narasimhan-Seshadri theorem.}.  This theorem states that
(under appropriate $\omega$-stability assumptions)
 there is a complex
gauge transformation making the associated
unitary connection self-dual \DoKro,
thus providing an identification of
$\CH\CB$ with $\CM^+ = \{{\nabla} \in \CA[E]: F^+=0\}/Aut(E)^{u} $,
where  $\CA[E]$ is
the space of unitary connections on $E$.
Physically, the relation is provided by
the quantum
equations of motion for $g$ in \holvbi:
 \eqn\eomi{
\omega\wedge  \biggl[ \nabla_{0}^{(0,1)} (g^{-1} \nabla_{0}^{(1,0)} g) -
\nabla_{0}^{(1,0)} \bar A
+ [g^{-1} \nabla_{0}^{(1,0)} g, \bar A]  \biggr] =0 \quad .
}
We view \eomi\ as
the $(1,1)$ part of the ASD Yang-Mills equation
$\hat F^+\CZ = 0 $.

In summary, we get the relation between the generating function
for current correlators and quantization of ${\CM}^{+}$.
Following the analogy to \wzwt\
we therefore make the following important
definition: {\it The space
$H^0(\CH\CB; \CL_\omega)$
is the space of (vacuum) 4D conformal blocks
for the  \wzwf\ model}.

{\bf Remark}.  We speculate that the conformal
blocks of nontrivial vertex operator correlators
associated to divisors can be obtained by quantization
of moduli spaces introduced by Kronheimer and Mrowka
\krmw. These are moduli spaces of ASD connections
with singularities along a divisor $D$ such that
the limit holonomy around $D$ exists and takes
values in a fixed conjugacy class.

\subsec{Projected Theory}

In this section we let  $X_{4}$ be an
algebraic surface, and denote it  as $S$.
Suppose we have a holomorphic map
\eqn\holmap{
f : S \to C
}
where $C$ is an algebraic curve.
Correspondingly, there are
maps of the moduli spaces, line bundles and
their cohomologies:

\eqn\mps{\eqalign{
& f^{*}: {\CH\CB}(C) \to {\CH\CB}(S) \cr
& (f^{*})^{*}: {\CL} \to L^{{\otimes} k} \cr
& (f^{*})^{*}: H^{i} ({\CH\CB}(S) , {\CL})
\to H^{i}({\CH\CB}(C), L^{{\otimes} k}) \cr}}

Here $L$ is the standard
determinant
line bundle over ${\CH\CB}(C)$,
and $k$ equals the integral of $\omega$ over a generic
fiber of $f$:
\eqn\cnchrge{
k = \int_{f^{-1}(x)} \omega
}

This simple observation can be interpreted in more physical terms
as follows:
Let us define the ``projected current'' to be:
\eqn\projcr{
J(x)dx \equiv \int_{f^{-1}(x)} \omega\wedge \gpg =
f_* ( \omega\wedge \gpg)
}
This
is a holomorphic 1-form on $C$ and
defines an ordinary two-dimensional current algebra
with  central charge
$k= f_{*}\omega$. Therefore we can
proceed to use all the standard constructions from
2D RCFT. For example, we can construct
 chiral vertex operators which  have all the familiar properties
of anomalous dimensions, monodromy representations
of the braid group, etc.

{\bf Remarks}

\item{1.}
In 2D RCFT chiral vertex operators are sometimes
heuristically regarded as path-ordered exponentials
of a chiral current. In the present context such an
identification would read as follows.
Let
$\Gamma$ be a curve in $C$ connecting
two points $x_{0}$ and $x_{1}$.
Then, given a representation $\rho_\lambda$ of
the group consider the formal relation:
\eqn\projvo{
\eqalign{
\rho_\lambda\Biggl(P\exp
	\int_\Gamma dx
\biggl\{
	\int_{f^{-1}(x)}
\omega\wedge \gpg\biggr\} \Biggr)
& = V_\lambda[D] V_{ \lambda^*}[D^{\prime}]\cr
D & = f^{-1}(x_{0}), D^{\prime} = f^{-1}(x_{1})\cr}
}
which
might be expected to define 4D chiral vertex operators.
In fact, it is a nontrivial problem to define a
proper regularization of this expression.
\foot{Similarly, the proper quantum definition of
a Wilson loop in a nonabelian gauge theory
is a highly nontrivial problem.}
{}From this point of view,  anomalous dimensions
arise because the projected Green function has
{\it logarithmic} singularities.

\item{2.}
It is natural to ask what kind of moduli space vertex operators
such as those discussed above might
correspond to.
In a  two dimensional theory the insertion of the vertex operator
at a point $x$ corresponds to the moduli space of holomorphic bundles
$\CE$
with parabolic
structure at the point $x$, i.e. there is a flag of subspaces in the
fiber
${\CE}\vert_{x}$
over the point.
The pullback of this bundle to $S$ will produce a holomorphic bundle
with
parabolic structure along the divisor $f^{-1}(x)$.
In this way we again relate correlation functions to the
quantization of Kronheimer-Mrowka moduli spaces,
since these may be identified with moduli spaces of
holomorphic vector bundles with parabolic structure
along a divisor \biquard\rade.

\newsec{\bf Related subjects: Hecke correspondences
and
Nakajima's algebras}

\subsec{Correlators and Correspondences}

The discussion of the previous section makes contact with
some
aspects of
the Beilinson-Drinfeld ``geometric Langlands program''
and also with some constructions of Nakajima.
Indeed, if
$h \pb_{\CE_1} h^{-1} = \pb_{\CE_2}$
then the holomorphic vector bundles are
isomorphic away from the singularities of $h$,
thus, under good conditions, they will
 fit into a sequence:
\eqn\exctseqi{
0\rightarrow
\CE_1 \rightarrow \CE_2 \rightarrow \CS\rightarrow 0
}
for some sheaf $\CS$.
Conversely, given $\CS$
we can form the
 ``geometric Hecke correspondence:''
\eqn\ghc{
\matrix{    &    &  \CP_\CS &  &  \cr
                & \pi_1 \swarrow &  &  \searrow \pi_2 \cr
       \CH\CB(c_1,c_2)& & & & \CH\CB(c_1',c_2')\cr}
}
defining, in a standard way,
 linear operators $(\pi_1)_* (\pi_2^*)$ and
$(\pi_2)_* (\pi_1^*)$ on  the
cohomology spaces. This construction
appears in the work of Beilinson and Drinfeld.
See also \nakajima\ginzburg\nakheis\ripoff.

One example of this construction is familiar in
CFT from the study 2D Weyl fermions \bost. The
correspondence analogous to \ghc\
defined by the sequences
\eqn\excseqiii{
\eqalign{
0 \rightarrow
\CL\otimes \CO(-P) \rightarrow & \CL \rightarrow \CL\mid_P
\rightarrow 0\cr
0 \rightarrow
\CL \rightarrow \CL\otimes \CO(Q) \rightarrow &
[\CL\otimes \CO(Q)]\mid_Q
\rightarrow 0\cr}
}
correspond to the effect of inserting $\psi(P), \bar \psi(P)$
respectively
in the correlation functions which define sections of
 $H^0(\CH\CB(c_1); \DET \pb_\CL) $.

It seems natural to conjecture that
if $h$ is singular along divisors or at points
then the operators $(\pi_1)_* (\pi_2)^*$ are equivalent
to insertion of the vertex operators
discussed in sections 4.3, 4.5.

\subsec{Nakajima's algebras}

In \nakajima\
Nakajima has shown that,  using the ADHM
construction of $U(k)$ gauge theory instantons
on ALE spaces $X_n=\widetilde{\IC^2/\IZ_n}$
one can construct highest weight representations
of affine Lie algebras.
More precisely, connections on $U(k)$ vector
bundles $E\to  X_n$ are specified by
$c_1(E)\in \Lambda_{root}(SU(n))$,
$\ch_2(E)\in \IZ_+$, and a flat connection at
infinity, i.e., a flat connection on the Lens
space $S^3/\IZ_n$. The McKay correspondence gives a 1-1
equivalence between flat $U(k)$ connections
on $S^3/\IZ_{n}$ and  integrable highest
weight representations of  $\widehat{SU(n)}_k$
at level $k$.  Let us denote this correspondence
as $\rho(\lambda) = \sum \ell_i \rho_i\in Hom(\IZ_n\to U(k)) /U(k)
\leftrightarrow
\lambda(\rho)\in HWT\bigl(\widehat{SU(n)}_k\bigr)
$
where $\ell_i$ are nonnegative integers assigned
to the nodes of the extended Dynkin diagram and
$\sum \ell_i = k$.

Nakajima's theorem states that
the cohomology of the moduli
spaces of  ``ASD instantons''
on $X_n$ is a representation
of $\widehat{SU(n)}_k$  at level $k$.
The reason for the quotation marks is explained
in the following section. Moreover,
$H^{p} ( \CM(c_1,\ch_2, \rho(\lambda)) )
$where $p = \half \dim \CM(c_1,\ch_2, \rho(\lambda)) $,
is naturally identified with the weight space
$\vec p = c_1,L_0= \ch_2(E)
$
of the representation $\lambda(\rho)$ of
$\widehat{SU(n)}_k$. Nakajima proves this
statement using the ADHM construction of
instantons. The generators of the affine Lie
algebra are defined using a geometric
Hecke correspondence, as described above.

\subsec{Torsion free sheaves}

We now address the question of exactly
which moduli space we should use in
Nakajima's theorem. \foot{This section represents work done
in collaboration with I. Grojnowski. See also \ripoff.
Related results have also been obtained by H. Nakajima
\nakheis. }
 The answer is {\it not} the moduli
space of instantons, as is clear by thinking
about the case $k=1$. Indeed, in this case we
are discussing
$U(1)$ gauge theory and
Nakajima's construction  gives a level one representation
of $\widehat{SU(n)}$.
But line bundles have no moduli !
Moreover, for line bundles
\eqn\lbdl{
\ch_2(\CL) = \half c_1(\CL)^2
}
or, in terms of CFT: $L_0 = \half \vec p^2$.
In terms of the Frenkel-Kac construction the
line bundles are states with  no
oscillators.

This problem is resolved when one realizes
that the generic
solution of the ADHM equations involves a generalization
of a line bundle. Indeed, the ADHM equations
can be solved rather explicitly
for the rank one case, and the moduli
space of ``line bundles'' on $X_n(\vec \zeta)$ turns out
to be:
\eqn\lbdlii{
\amalg_{\ell\geq 0} X_n^{[\ell]}  \times
H^2(X;\IZ)
}
where $X_n^{[\ell]} $ is the Hilbert scheme of
$\ell$ points on $X_n$.
The proper interpretation of this moduli space
is that it is the space of
{\it ``torsion free sheaves.''} These are sheaves
 which fit in an
exact sequence:
$0 \rightarrow\CE \rightarrow \CL  \rightarrow
\CS  \rightarrow 0 $
where
$\CS$ is a coherent sheaf supported at points.
\foot{They are called ``torsion free'' because,
as $\CO$ modules they have no torsion.}
In the simplest case where $\CS$ is a
skyscraper sheaf at  $x=y=0$,   $\CE$ is given
by the  $\IC[x,y]$ module consisting of
functions {\it vanishing}  at $x=y=0$. In sharp
contrast to the situation in $2D$, $\CE$ cannot
be considered to be a line bundle.

Once we have admitted the necessity for
torsion free sheaves we may easily evaluate
the ``quantum numbers''  of $\CE$ using the Riemann-Roch-
Grothendieck theorem:
\eqn\lbiii{
\eqalign{
c_1(\CE) & = c_1(\CL)\cr
  -\ch_2(\CE)& = - \half (c_1(\CL))^2 + \ch_2(\CS)
= \half \vec p^2 + \ell \cr}
}
Here $\ell$ is, generically, the number of points
at which $\CS$ has its support.
Thus we recover
$L_0 = \half \vec p^2 + \ell $.

Another immediate benefit of admitting
torsion free sheaves is that we can
reproduce the Frenkel-Kac characters from
Nakajima's approach. In particular the
character is:
\eqn\fkchar{
{\Tr}_\CH q^{-\ch_2(\CE)} =
\sum q^{-k(\CE)} \dim H^{\half}(X_\zeta^{[\ell]})
 =\sum_{p\in \Lambda_{root} } {q^{\half p^2}\over  \eta^r}
}
where $H^{\half}$ is the middle-degree
cohomology.
The reader will note the similarity to
the calculation of \VaWi, which applies the
results of  \gottsh\hirz\gothuy\  to evaluate  the $N=4$ SYM
partition function on $K3$. In general it appears
that the compactification by torsion free sheaves,
also known as the Gieseker compactification, is
the one most relevant to physical theories.

\newsec{Five-Dimensional Viewpoint.
Verlinde Formula and Donaldson Invariants}

When $X_5= X_4 \times \IR$, $X_4$ K\"ahler, we
may define the
5D ``K\"ahler-Chern-Simons''
theory:
\eqn\kcs{
\eqalign{
S =
\int_{X^{4} \times S^{1}}  & Tr ( {\omega} \wedge
[ A dA + {2\over{3}} A^{3} + dt \wedge \psi \wedge \bpsi] + \cr
& + dt \wedge [\hi \wedge \pba \bpsi + \bar\bhi \wedge \pa \psi +
\hh F^{0,2} + \bar\bh F^{0,2} ])\cr}
}
Here $A$ is a 5D gauge field, while
$\bar H^{0,2}\in \Omega^{0,2}(X_4,\lieg)$ etc. are
bosonic and $\chi^{2,0}\in \Omega^{2,0}(X_4,\lieg)$,
$\psi\in \Omega^{1,0}(X_4,\lieg)$ are fermionic.
The quantization associated with the
first line of \kcs\ is straightforward and we skip the details.
Consequently, on $X_5=X_4\times S^1$ we have
the partition function
$Z_{KCS} = \dim H^0(\CH\CB; \CL_\omega)$.

{\bf Remarks}

\item{1.} The Lagrangian (without the fermions)
has appeared before in the work of
Nair and Schiff \ns\nair. It is necessary to
introduce the Lagrange multipliers
 $H, \bar H$ for  constraints restricting the
integral to an integral over $\CA^{1,1}$.
The constraints $F^{2,0} = F^{0,2} = 0$
are second class,  necessitating
the fermions. Indeed, $\CA^{1,1}$
 is invariant under complexified gauge
transformations $\CG_c$
and $\bar H^{0,2}$
fixes this  invariance. The gauge
fixing introduces  the
extra terms involving $\chi,\psi$.
Put another way,  we are simply
 writing the  Poincar\'e
dual,
$\eta(\CA^{1,1}\hookrightarrow \CA) $
using a Mathai-Quillen representation.
Following the standard procedure one  introduces
the topological multiplet $(A,\psi)$ and the
antighost multiplets:
$(\chi^{2,0}, H^{2,0}) , (\bar \chi^{0,2}, \bar H^{0,2})$.
\foot{For a review of this
technology see \CMR. } The new point here is that there is
an extended symmetry in the problem (time-dependent
gauge transformations, extended by the $U(1)$, rotating
the time loop), and the Poincare dual as well as the whole
action is equivariantly closed with respect to
this  extended symmetry.

\item{2.} In the systematic development of 2D RCFT
from the 3D CSW point of view
one derives the chiral current  algebra and representations
of \wzwt\ from quantization of CSW on a 3-manifold
with boundary, e.g. on $B_2\times \IR$
\wittjones\taming\elitzur.
In the 5D KCS/\wzwf\  theory there are
new complications. Nevertheless, similar
manipulations allow one to recover
the algebra $\kappa(X_3,\lieg,\omega)$
we found in section 2.4  as follows.
We let the $4$-fold have a boundary,
$\p X_4 = X_3\not=\emptyset$, e.g.,
$X_3 = S^3$ or $S^3 \amalg S^3$,
and we take the gauge group to be:
$
\CG_1=\{g(x)\in Aut(P): g\vert_{X_3} =1\}
$. This group
acts symplectically on $\CA$ with moment map
\eqn\mommp{
\mu(\epsilon) =
\int_{X_4} \omega {\Tr} (\epsilon F) + \int_{X_3} {\Tr}(\epsilon
\omega
\wedge A)
}
The analog of the
moduli space of instantons is the infinite-dimensional
symplectic quotient:
$
\mu^{-1}(0)\cap \CA^{1,1} /\CG_c \equiv \CM^+
$.
The current algebra is obtained from the
algebra of moment maps $\mu(\epsilon)$
and  is just $\kappa(X_3,\lieg,\omega)$.

\subsec{The Verlinde Formula}

The 4D version of the Verlinde formula
can be derived by localizing the 5D KCS
path integral on the 5-manifold $X_4 \times S^1$,
viewing the latter as an example of integration
of equivariant differential forms and
using a BRST symmetry $Q$. The calculation is
closely related to that of
\gerasimov\BlThlgt, as well as the recent calculations
of \parki. As in \parki\  there are two branches of fixed points
called A and B. We present here a  {\it preliminary}
answer for the contribution of branch B.
Deriving the result for branch A requires further
techniques which are under investigation \fdrcft.

A fairly extensive calculation, for
$\CE$ of rank $r$ and $c_1(\CE)=0$,
expanding around the
fixed points of $Q$ and evaluating the determinants
of quadratic fluctuations leads to a formula given by
a sum over decomposition into line bundles (``abelianization'')
$\CE \cong \oplus \CL_i$. When $b_1(X) =0$
and $p_g(X) = h^{2,0}(X) =0$
the formula specializes to:

\def\fcx{c_{1}(X)}
\def\scx{c_{2}(X)}
\eqn\answri{\eqalign{
\dim H^0(\CH\CB,\CL_\omega) =
\sum_{\zeta \in {\rm Pic}(X^{4})^{r}, [\zeta] \cdot [\zeta] = -k }
 &
\int_{\liet \cap \Delta_{+}} d \phi
e^{ - i\half
\langle \phi , [\zeta] \rangle
 \cdot ( \omega + h \fcx )}
  \cr
 (-1)^{\langle \rho, [\zeta] \rangle \cdot c_1(X) } \times
{\prod_{\alpha>0}} (
2 \sin({ \phi_{\alpha} \over{2}})
&
)^{
 [\zeta]_{\alpha} \cdot [\zeta]_{\alpha}  +
{1\over{6}}(\fcx^{2} + \scx)}\cr
}
}
where
$[\zeta] \in H^{2}(X_{4}, {\IZ}^{r})$ is the
collection of first Chern classes of $\zeta$,
$a\cdot b = \int_{X_{4}} a \wedge b$.
With the identification of $\IR^r$ as the
Cartan subalgebra of $SU(r+1)$ we have
$\phi_\alpha =\langle \phi, {\alpha}\rangle$,
for the roots ${\alpha}$,
$\rho = {\half} \sum_{\alpha > 0} {\alpha}$,
$h$ is the dual Coxeter number and
$\liet \cap \Delta_{+}$ is the Weyl alcove.
For  $SU(2)$   \answri\
 may be further simplified to
\eqn\sutwo{
\int_0^{2 \pi} d \phi \biggl({1 \over 2 \sin \phi/2} \biggr)^{4k-2}
\CP(X;k,H,\phi)
}
where $\CP(X;k,H,\phi)$ is a manifold-dependent
distribution
\eqn\dstrb{
\CP(X;k,H, \phi) =
\sum_{\zeta \in {\rm Pic} (X^{4}) , [\zeta] \cdot [\zeta] = -k }
(-1)^{[\zeta]\cdot c_1(X)} e^{-i\half \phi  [\zeta]\cdot (H+2 c_1(X))
}
}
with
$H$ the divisor corresponding to the K\"ahler class.

When $h^{2,0}>0$ the existence of fermion zeromodes
leads to a more compliciated version
of \answri\  involving a Grassmann integral over
$H^{2,0}(X)\otimes\liet $. The derivation of \answri\
and its generalization for $h^{2,0}>0$ will be
presented in \fdrcft.

{\bf Remark}.
 As usual the integral \sutwo\ is ill-defined because
of endpoint divergences. By analogy to previous
discussions of nonabelian localization the sense
of the contour deformation should be dictated by
the sign of the moment map
\ref\kirwan{L. Jeffrey and F. Kirwan,
``Localization for nonabelian group actions'',   alg-geom/
9307001}
\park. In the resulting formula,
 for some choices
of manifold $X_4$, second Chern character
 $k$, and embedding $H$,
 \sutwo\ then leads
to well-defined integers, even without consideration of
the $A$-branch.

\subsec{Semiclassical Limit}

A semiclassical limit is of particular interest because
it leads to a very computable theory.
We set $\omega = k \omega_0$, for some
fixed K\"ahler class $\omega_0$ and
take  $k\to \infty$.
By the  index
theorem $\dim H^0 = \int \ch(\CL) Td(\CH \CB) $
\foot{assuming higher cohomology groups are zero}
so in the large $k$ limit we get:
\eqn\lmtone{
Z_\omega^{KCS}\to
k^{\dim \CM^+/2} \Vol_{\omega_0}(\CM^+)
}

On the other hand,
taking the limit in the KCS
path integral we localize to the space of
time-independent fields and get:
\eqn\limpthint{
\eqalign{Z_\omega^{KCS}\to k^{\dim \CM^+/2}
\int & {[d A d\psi  d \chi d H d \phi ] \over  \vol \CG}  \cr
\exp\Biggl\{
-{i \over  4 \pi}
\int_{X_4}  {\Tr}\biggl[
\bar H^{0,2} F^{2,0} + \phi \omega_0\wedge F^{1,1} &
 + H^{2,0} F^{0,2}\cr
+
\omega_0 \psi\wedge \bar \psi +   \bar \chi^{0,2}  &{\p}_{A} \psi +
 \chi^{2,0} \pb_{A} \bar \psi \biggr] \Biggr\}  \cr}
}
This theory has been considered before in
\park. The authors of \park\ refer
to the theory as  ``holomorphic Yang-Mills
theory,'' (HYM).  They
derived it in order to produce a
QFT expression for the symplectic volume
of moduli space.  Hyun and Park
derived a  formula analogous
to \answri\  from a localization argument
in \parki.

{\bf Remarks.}

\item{1.}  The volume of the instanton moduli space is one of the
Donaldson
observables,
\eqn\vlm{\langle \exp ( \int_{X_{4}} \omega \wedge
{\Tr} {\phi} F + {\half} {\psi} {\psi} ) \rangle}
and for a large class of K\"ahler surfaces this was computed recently
in \Witfeb.

\item{2.} On hyperk\"ahler manifolds one can integrate out
$\psi$ to produce a Lorentz invariant action.
Moreover, using the ADHM construction it is
possible to write fairly explicit results for the
path integral \limpthint.

\centerline{\bf Acknowledgements}

We would like to thank
A. Alekseev, A. Beilinson, R. Dijkgraaf,
M. Douglas, L.Faddeev, I. Frenkel, D. Friedan,
A. Gerasimov,
I. Grojnowski, D. Gross, J. Harvey, A. Johansen,
B. Khesin,
M. Kontsevich,
A. Levin,
D. Morrison, H. Nakajima, P. van Nieuwenhuizen,
A. Polyakov,
A. Rosly, S. Shenker,
I. Singer, W. Taylor, K. Uhlenbeck,
and A. Zamolodchikov
for useful remarks and discussions.
S.S. would like to thank the Theory Division of
CERN and G.M would like to thank
the Aspen Center for Physics  for hospitality.
The research of G. Moore
is supported by DOE grant DE-FG02-92ER40704,
and by a Presidential Young Investigator Award; that of S.
Shatashvili,
by DOE grant DE-FG02-92ER40704, by NSF CAREER award and by
OJI award from DOE.

\listrefs
\bye